\documentclass[sn-nature]{sn-jnl}

\usepackage{multirow}%
\usepackage{amsmath,amssymb,amsfonts}%
\usepackage{amsthm}%
\usepackage{mathrsfs}%
\usepackage[title]{appendix}%
\usepackage{xcolor}%
\usepackage{textcomp}%
\usepackage{manyfoot}%
\usepackage{booktabs}%
\usepackage{algorithm}%
\usepackage{algorithmicx}%
\usepackage{algpseudocode}%
\usepackage{listings}%

\usepackage{microtype}
\usepackage{graphicx}
\graphicspath{{./figures/}}

\usepackage{leftindex}
\usepackage{amsmath,amssymb,amsfonts}
\usepackage{bm}
\usepackage[capitalize]{cleveref}
\usepackage{xspace}
\usepackage{dsfont}
\usepackage{booktabs}
\usepackage{mathtools}
\usepackage{colortbl}
\usepackage{placeins}
\usepackage{braket}
\usepackage{soul}
\usepackage{cancel}
\usepackage{lineno}

\unnumbered

\begin{document}

\title[Statics and Dynamics of non-Hermitian Many-Body Localization]{Statics and Dynamics of non-Hermitian Many-Body Localization}


\author*[1]{\fnm{J\'ozsef} \sur{M\'ak}}\email{jozsef.mak@kcl.ac.uk}

\author[1]{\fnm{M. J.} \sur{Bhaseen}}%

\author[2]{\fnm{Arijeet} \sur{Pal}}%

\affil[1]{\orgdiv{Department of Physics}, \orgname{King's College London}, \orgaddress{\street{Strand}, \city{London}, \postcode{WC2R 2LS}, \country{United Kingdom}}}

\affil[2]{\orgdiv{Department of Physics and Astronomy}, \orgname{University College London}, \orgaddress{\street{Gower Street}, \city{London}, \postcode{WC1E 6BT}, \country{United Kingdom}}}

\abstract{
Many-body localized phases retain memory of their initial conditions in disordered interacting systems with unitary dynamics. The stability of the localized phase due to the breakdown of unitarity is of relevance to experiment in the presence of dissipation. Here we investigate the impact of non-Hermitian perturbations on many-body localization. We focus on the interacting Hatano-Nelson model which breaks unitarity \textit{via} asymmetric hopping. We explore the phase diagram for the mid-spectrum eigenstates as a function of the interaction strength and the non-Hermiticity. In contrast to the non-interacting case, our findings are consistent with a two-step approach to the localized regime. We also study the dynamics of the particle imbalance. We show that the distribution of relaxation time scales differs qualitatively between the localized and ergodic phases. Our findings suggest the possibility of an intermediate dynamical regime in disordered open systems.
}

\maketitle

\section*{Introduction}\label{sec1}

The investigation of isolated quantum systems has led to a remarkable understanding of novel states of matter \cite{bloch2008many,eisert2015quantum}. However, naturally occurring and engineered quantum systems are typically coupled to an environment, even if the coupling is weak. The dynamics of such open systems, over a reduced set of trajectories, can often be described by an effective non-Hermitian Hamiltonian which breaks unitarity \cite{daley2014quantum}. These have been realized in pioneering experiments on photonic \cite{ruter2010observation,regensburger2012parity,feng2017non,longhi2018parity} and matter-light \cite{zhang2016observation,peng2016anti,zhang2018non,li2019observation,ozturk2021observation} systems. The study of non-Hermitian Hamiltonians allows an enriched classification scheme for describing quantum matter \cite{el2018non,kawabata2019symmetry}. For example, non-Hermitian systems with parity and time-reversal symmetry can have real eigenvalues, much like their Hermitian counterparts \cite{bender2002generalized,bender2010pt,brody2013biorthogonal}. Non-Hermitian perturbations can also lead to novel phases and phase transitions beyond the equilibrium paradigm \cite{ozturk2021observation,sergeev2021new,fruchart2021non,weidemann2022topological}. For a review of the applications of non-Hermitian systems see \cite{ashida2020non}.

An important class of isolated quantum systems are those that fail to equilibrate on long timescales. This includes many-body localized (MBL) phases which retain memory of their initial conditions, as observed in analog and digital quantum simulators \cite{schreiber2015observation,smith2016many,xu2018emulating,lukin2019probing,guo2020observation,smith2019simulating,zhu2021probing}. The fate of MBL in open systems has also been investigated in cold atom settings via a controlled coupling to the environment \cite{vznidarivc2015relaxation,levi2016robustness,fischer2016dynamics,medvedyeva2016influence,bordia2016coupling,luschen2017signatures,rubio2019many}. This is of considerable interest in solid state devices due to their intrinsic coupling to other degrees of freedom \cite{abanin2019colloquium,nietner2022route}. The stability of MBL has also been studied \cite{de2017stability,de2017many} including the effects of local dissipation in the Lindblad formalism \cite{vznidarivc2015relaxation,levi2016robustness,fischer2016dynamics,medvedyeva2016influence,everest2017role,wybo2020entanglement,morningstar2022avalanches,sels2022bath}, and by coupling to delocalized environments; see for example \cite{rubio2019many,kelly2020exploring}. Effective non-Hermitian models can also provide insight into the stability of quantum matter in the presence of coupling to an environment. In the context of non-Hermitian MBL, pioneering studies have focused on the link between the phase diagram and spectral properties \cite{hamazaki2019non}, together with mobility edges \cite{Heussen2021,panda2020entanglement}. Extensions of these investigations have also considered wavepacket dynamics \cite{orito2022unusual} and the effects of quasiperiodic potentials \cite{zhai2020many,suthar2022non}. 

\begin{figure}[tp!]
	\centering
	\includegraphics[width=3.2in]{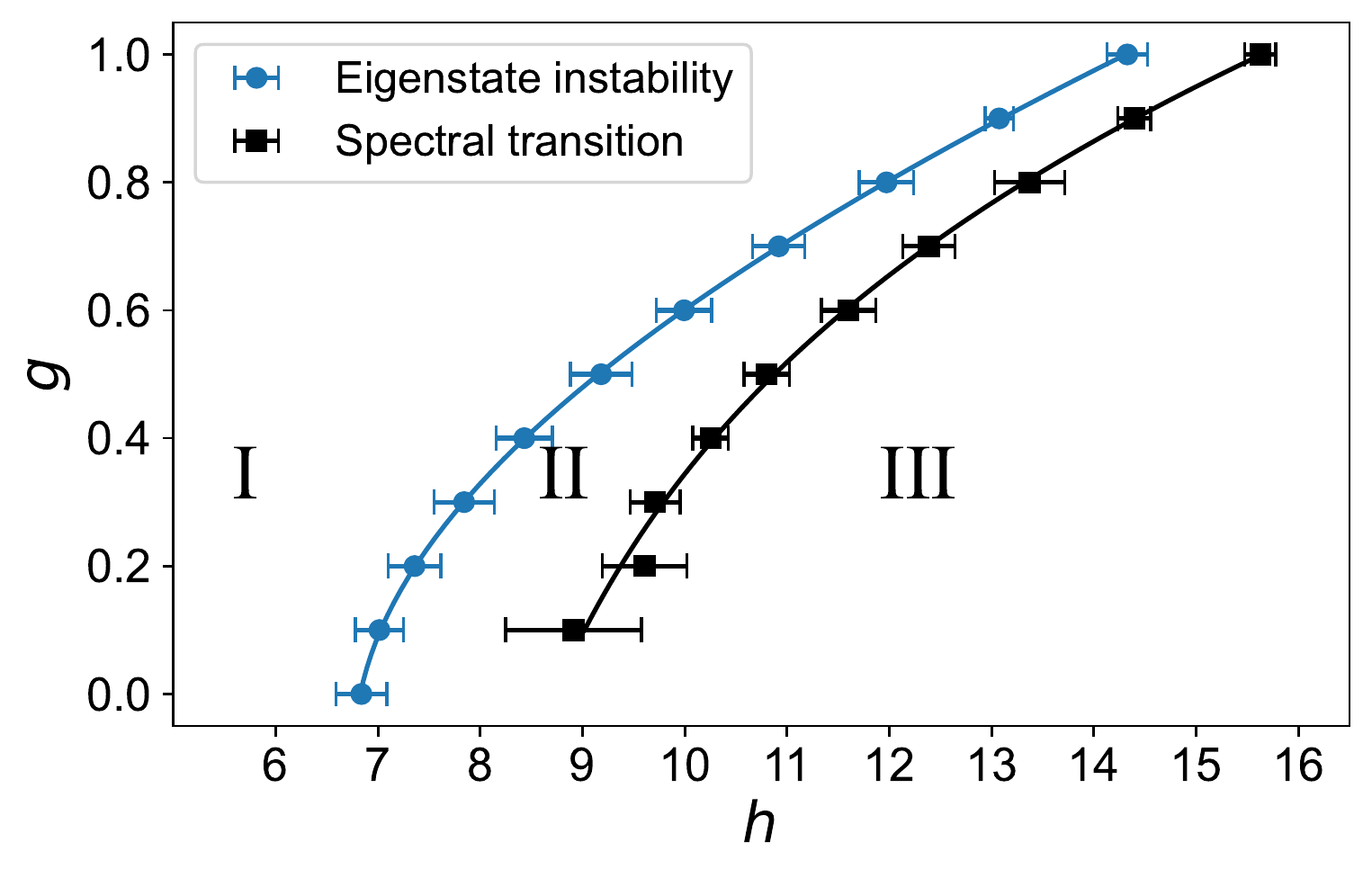}
	\caption{
		\textbf{Phase diagram of the interacting Hatano-Nelson model as a function of the non-Hermiticity parameter $g$ and the disorder strength $h$.} We set the hopping strength to $J=1$ and the interaction strength to $U=2$. The results are obtained using exact diagonalization (ED) with $L=8,10,12,14,16$ sites. The left boundary (blue line) corresponds to the instability $\mathcal{G}$ of real eigenstates to a non-Hermitian perturbation. The right boundary (black line) indicates the spectral feature where the fraction of complex energy eigenvalues $f_\mathrm{C}$ goes from increasing to decreasing with increasing $L$. Region I corresponds to unstable real eigenstates and increasing $f_\mathrm{C}$. Region II corresponds to stable real eigenstates and increasing $f_\mathrm{C}$. Region III corresponds to stable real eigenstates and decreasing $f_\mathrm{C}$. The error bars indicate the variation of the finite-size crossing points for each data set.
	}
	\label{fig:g-vs-h}
\end{figure}

In this work we explore the phase diagram of the interacting Hatano-Nelson model as a function of non-Hermiticity and the interaction strength; see Fig. \ref{fig:g-vs-h}. Using exact diagonalization, we provide evidence for a two-step approach towards the localized regime, with a significant separation between eigenstate and spectral transitions. The latter yields a dynamical instability within the localized regime. We provide predictions for the relaxation time scales of the particle imbalance with a view towards cold atom experiments.

\section*{Results}

\subsection*{Model}

We consider an interacting version of the one-dimensional Hatano-Nelson model \cite{hatano1996localization,hatano1997vortex,hatano1998non,brouwer1997theory,hatano1998localization} with $L$ sites and periodic boundary conditions
\begin{equation}
	\label{eq:mbl-nh-model}
	\hat{H}=\sum_{i=1}^{L}\left[-J\left(\mathrm{e}^{-g} \hat{b}_{i} \hat{b}_{i+1}^{\dagger} + \mathrm{e}^{g} \hat{b}_{i}^{\dagger} \hat{b}_{i+1}\right)+\\
	U \hat{n}_{i} \hat{n}_{i+1}+h_{i} \hat{n}_{i}\right],
\end{equation}
where $J$ is the hopping strength, $g$ parametrizes the hopping asymmetry, $h_i$ represents on-site disorder and $U$ is the nearest-neighbor interaction strength. For simplicity we consider hard-core bosons $\hat{b}_i$ and $\hat{b}_i^\dagger$ where $\hat{n}_i=\hat{b}_i^\dagger\hat{b}_i$. The disorder is drawn from a uniform distribution $h_i \in [-h,h]$. The asymmetry in the hopping terms renders the model non-Hermitian when $g\neq0$. Throughout this work we consider half-filling and set $J=1$.

As discussed in Supplementary Note 1, the model \eqref{eq:mbl-nh-model} can be derived within the Lindblad formalism by assuming a specific form of the jump operators \cite{gong2018topological} and post-selecting on jump-free trajectories. It therefore provides a useful framework for investigating the stability of MBL when coupled to an environment. Although the experimental realization of such systems is hindered by the need for post-selection, this may be within reach for modest system sizes \cite{naghiloo2019quantum,wu2019observation}. In particular, non-reciprocal transport has been realized in cold atoms \cite{liang2022dynamic} and non-Hermitian effects have been simulated with trapped ions \cite{noel2022measurement}. Refs \cite{gong2018topological,mu2020emergent,lee2014heralded} discuss the experimental realizations of non-Hermitian models related to the model \eqref{eq:mbl-nh-model} using cold atoms.

Before embarking on a detailed study of the model (\ref{eq:mbl-nh-model}) we first discuss some limiting cases. In the non-interacting limit with $U=0$, all states are localized for $g=0$ \cite{abrahams201050}. However, for $g\neq0$ it undergoes a delocalization transition \cite{hatano1996localization,hatano1997vortex,hatano1998non,brouwer1997theory,hatano1998localization}. In the Hermitian case ($g=0$) with $U\neq0$, the model exhibits a many-body localization (MBL) transition \cite{pal2010many,serbyn2015criterion,nandkishore2015many,alet2018many,abanin2019colloquium}. In the clean limit ($h=0$) the model exhibits $\mathcal{PT}$-symmetry breaking transitions in both the ground state and excited state sectors with only the former surviving to the thermodynamic limit \cite{zhang2022symmetry}. In this paper we study the interplay between non-Hermiticity and MBL by considering both the mid-spectrum states of the model (\ref{eq:mbl-nh-model}) and its dynamics. To obtain the mid-spectrum states we employ exact diagonalization (ED) and retain a total of $N_\mathrm{T}=\lceil 0.04\mathcal{N} \rceil$ eigenstates (rounded up to the nearest integer) which are closest to the mid-point of the spectrum, $\mathrm{Tr}(H)/\mathcal{N}$; here $\mathcal{N}=\binom{L}{L/2} \sim 2^L/\sqrt{L}$ is the dimension of the Hilbert space.

\subsection*{Symmetry}

\begin{figure}[t]
	\centering
	\includegraphics[width=3.2in]{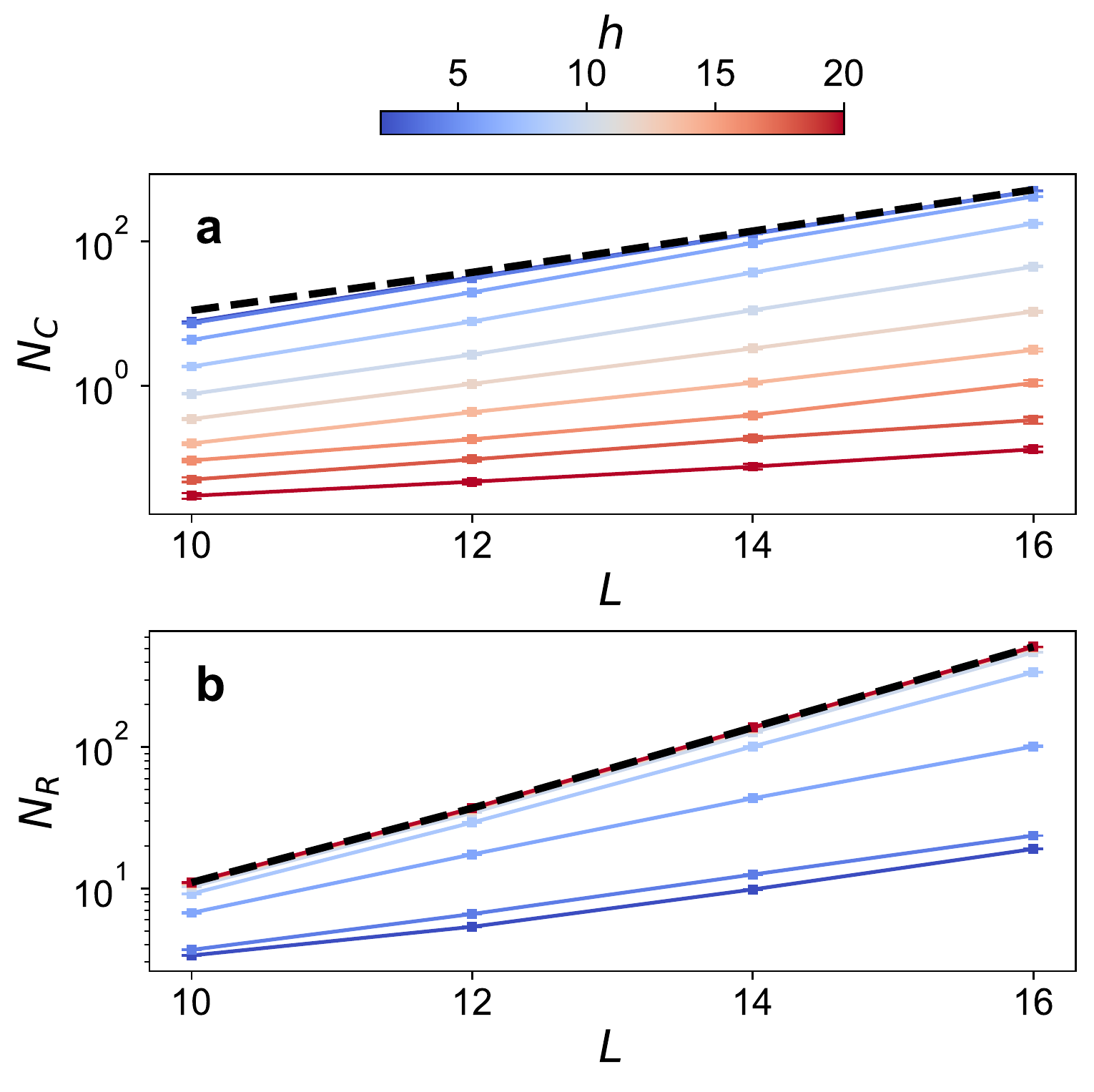}
	\caption{
		\textbf{The number of real and complex eigenvalues in the interacting Hatano-Nelson model as a function of the disorder strength $h$.} \textbf{a} The number of complex eigenvalues $N_\mathrm{C}$ and \textbf{b} real eigenvalues $N_\mathrm{R}$ both grow exponentially as a function of the system size $L$. We set the hopping strength to $J=1$, the interaction strength to $U=2$ and the non-Hermiticity parameter to $g=0.5$. The black dashed line corresponds to the total number of computed eigenpairs $N_\mathrm{T} \sim 2^L/\sqrt{L}$. The data points correspond to the mean over $10^4$ disorder realisations. The error bars correspond to the standard error of the mean; for weak disorder they are smaller than the markers.
	}
	\label{fig:eigenvalues}
\end{figure}

The non-Hermitian Hamiltonian (\ref{eq:mbl-nh-model}) exhibits time-reversal symmetry which ensures that the eigenvalues are either real or occur in complex conjugate pairs \cite{hamazaki2019non,hamazaki2020universality}. In the non-interacting limit with $U=0$ it has been shown that the localized eigenstates correspond to purely real eigenvalues and the delocalized states correspond to complex conjugate pairs \cite{hatano1996localization,hatano1997vortex,hatano1998localization,hatano1998non,brouwer1997theory}. At large disorder all states are localized for $U=0$, corresponding to an entirely real spectrum. In contrast, in the interacting case it is only the fraction of complex eigenvalues that goes to zero. In fact, the number of complex eigenvalues remains non-zero and grows exponentially with increasing system size, as shown in Fig.~\ref{fig:eigenvalues}. For weak disorder the number of complex eigenvalues $N_\mathrm{C}$ approaches the total number of computed eigenvalues $N_\mathrm{T} \sim 2^L/\sqrt{L}$ but the number of real eigenvalues $N_\mathrm{R}$ remains non-zero, see Fig.~\ref{fig:eigenvalues}(a). Conversely, in the strong disorder regime the number of real eigenvalues approaches the total number of eigenvalues but the number of complex eigenvalues remains non-zero, see Fig.~\ref{fig:eigenvalues}(b). As we discuss more fully below, in the many-body problem localized eigenstates may correspond to complex eigenvalues.

\subsection*{Spectral Transition}

\begin{figure}[t]
	\centering
	\includegraphics[width=3.2in]{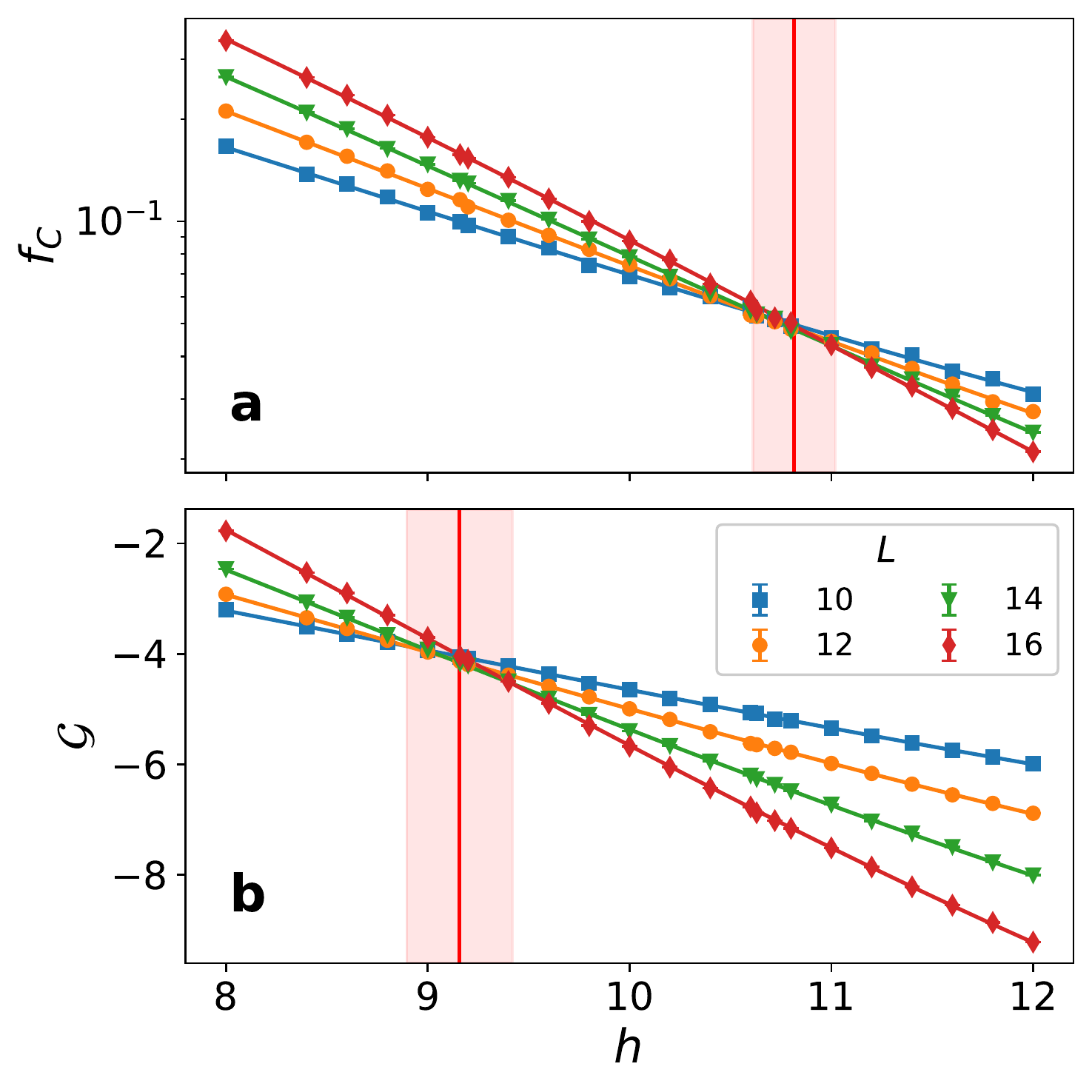}
	\caption{
		\textbf{Determination of the locations of the spectral and eigenstate transitions.} \textbf{a} Variation of the fraction of complex eigenvalues $f_\mathrm{C}$ as a function of the disorder strength $h$ for different system sizes $L$, with the non-Hermiticity parameter $g=0.5$ and the interaction strength $U=2$. The crossing point at $h \approx 10.8$ shows the separatrix between regions II and III in Fig.~\ref{fig:g-vs-h}. An eigenvalue is considered real if its absolute imaginary part is below $10^{-13}$ and complex otherwise. \textbf{b} Variation of the eigenstate instability measure $\mathcal{G}$ as a function of the disorder strength $h$ for different system sizes $L$, with $g=0.5$ and $U=2$. The crossing point at $h \approx 9.1$ shows the separatrix between regions I and II in Fig.~\ref{fig:g-vs-h}. In our finite-size simulations the transitions in panels \textbf{a} and \textbf{b} are well-separated. The vertical shaded regions extend between the lowest and highest values of $h$ where the data for different values of $L$ cross. The vertical line (red) indicates the mid-point of the region. The mid-point and region width are plotted as the markers and the error bars, respectively, in Fig. \ref{fig:g-vs-h}. In both panels the data points are computed as the mean over $10^4$ disorder realizations. The error bars corresponding to the standard error of the mean are smaller than the markers.
	}
	\label{fig:G-fC-scaling}
\end{figure}

Following Ref. \cite{hamazaki2019non} we first consider the behavior of the fraction of complex eigenvalues $f_\mathrm{C} = \overline{N_\mathrm{C}/N_\mathrm{T}}$ with increasing disorder strength, where the overbar denotes disorder averaging. As shown in Fig.~\ref{fig:G-fC-scaling}(a) the spectrum of the Hamiltonian undergoes a transition at a critical disorder strength where $f_\mathrm{C}$ goes from increasing to decreasing with increasing system size \cite{hamazaki2019non}. However, the number of complex eigenvalues $N_\mathrm{C}$ does not go to zero with increasing $L$ in the strong disorder regime. This is in contrast to the non-interacting ($U=0$) case \cite{brouwer1997theory}. In Fig.~\ref{fig:g-vs-h} we plot the evolution of this boundary as a function of the non-Hermiticity $g$ and the disorder strength $h$. It can be seen that at large non-Hermiticity the transition occurs for larger disorder strength.

\subsection*{Eigenstate Transition}

Having established the locus of the spectral transition, we now turn our attention to the stability of the eigenstates. In the Hermitian case, one of the hallmarks of localized eigenstates is their stability to local perturbations \cite{serbyn2015criterion}. An extension of this to the non-Hermitian case was suggested in Ref. \cite{hamazaki2019non}. First, we denote the left and right eigenstates of the non-Hermitian Hamiltonian (\ref{eq:mbl-nh-model}) by $\ket{E_k}_\mathrm{R}$ and $\ket{E_k}_\mathrm{L}$. These satisfy $\hat{H}\ket{E_k}_\mathrm{R}=E_k\ket{E_k}_\mathrm{R}$ and $\leftindex_{\mathrm{L}}{\bra{E_k}}\hat{H}= \leftindex_{\mathrm{L}}{\bra{E_k}} E_k$ with the complex eigenenergies $E_k=\mathcal{E}_k+\mathrm{i}\Lambda_k$. Denoting the eigenstates which correspond to purely real eigenenergies by $\ket{\mathcal{E}_k}_\mathrm{R}$ and $\ket{\mathcal{E}_k}_\mathrm{L}$ we examine the quantity
\begin{equation}
\label{eq:stability-def}
\mathcal{G}(h)=\overline{\ln \left| \frac{\leftindex_{\mathrm{L}}{\braket{\mathcal{E}_{k+1}|\hat{V}_{NH}|\mathcal{E}_k}}_\mathrm{R}}{\mathcal{E}_{k+1}'-\mathcal{E}_k'}\right|},
\end{equation}
where $\hat{V}_{NH}$ is a non-Hermitian perturbation and $\mathcal{E}_k'=\mathcal{E}_k+\leftindex_{\mathrm{L}}{\braket{\mathcal{E}_k|\hat{V}_{NH}|\mathcal{E}_k}}_\mathrm{R}$ is the perturbed eigenvalue. Here we take $\hat{V}_{NH}=\hat{b}_1^\dagger\hat{b}_2$ and the eigenstates are ordered with increasing $\mathcal{E}'_k$. In Fig.~\ref{fig:G-fC-scaling}(b) we plot the evolution of the instability parameter $\mathcal{G}$ versus disorder strength for different system sizes. It can be seen that the eigenstates are stable (unstable) above (below) a critical disorder strength. However, the value of the critical disorder strength differs from that obtained from $f_\mathrm{C}$. In Fig.~\ref{fig:g-vs-h} we plot the locus of the stability line obtained from Eq. (\ref{eq:stability-def}). It can be seen to be well-separated from the spectral transition in our finite-size simulations. That is to say, the instability of the real eigenstates occurs in advance of the spectral transition where $f_\mathrm{C} \to 0$ with increasing $L$. This suggests the possibility of a two-step transition to the localized phase in open systems. This is consistent with results obtained from the study of mobility edges \cite{panda2020entanglement}.

\subsection*{Role of Interactions}

\begin{figure}[t]
	\centering
	\includegraphics[width=3.2in]{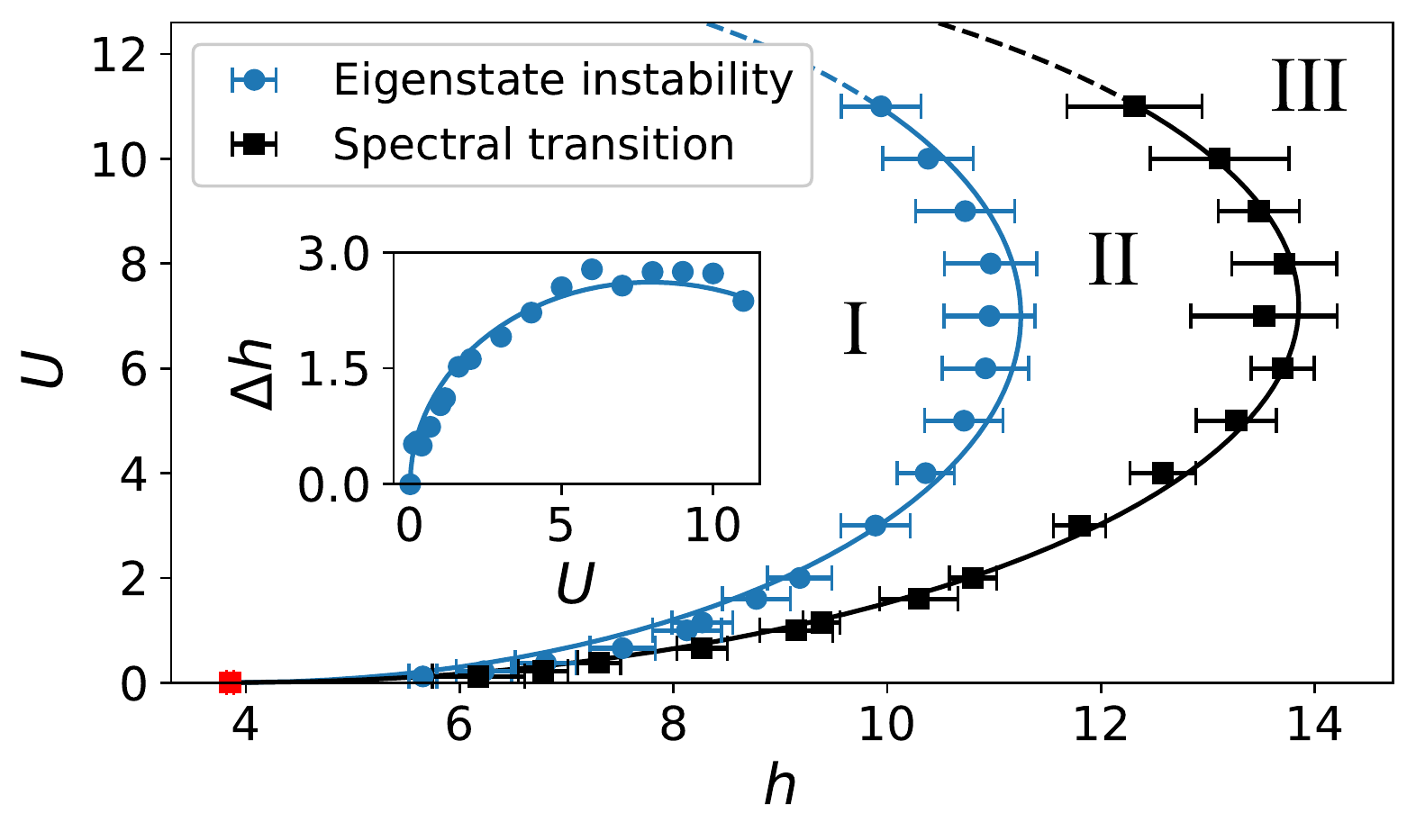}
	\caption{
		\textbf{Evolution of the phase diagram of the interacting Hatano-Nelson model as a function of interaction strength $U$ and the disorder strength $h$.} We set the non-Hermiticity parameter to $g=0.5$ and the hopping strength to $J=1$. The transition points and error bars are extracted using the method described in Fig.~\ref{fig:G-fC-scaling} using the same system sizes. Regions I-III are defined in Fig.~\ref{fig:g-vs-h}. The single-particle localization transition (red square) is computed from the crossing points of the fraction of complex eigenvalues $f_\mathrm{C}$ for $L=100,200,300$. Inset: evolution of the width $\Delta h$ of region II, corresponding to the horizontal separation between the data points in the main panel, as a function of $U$.
	}
	\label{fig:u-vs-h}
\end{figure}

Having established the presence of two boundaries in the non-Hermitian problem we now turn our attention to the role of the interaction strength. In Fig.~\ref{fig:u-vs-h} we show the evolution of the boundaries with increasing $U$ for a fixed value of the non-Hermiticity $g$. It can be seen that in the non-interacting limit with $U=0$ the transitions coincide \cite{hatano1997vortex,brouwer1997theory,kawabata2021nonunitary}; that is to say the localization transition coincides with the spectral transition from complex to fully real. However, in the presence of interactions the phase structure changes. The two transitions separate with increasing $U$. This suggests the possibility of an intermediate regime as one goes from weak disorder to strong, as found in Fig.~\ref{fig:g-vs-h}.

\subsection*{Dynamics}

To further characterize the localized and delocalized phases, we turn our attention to non-equilibrium dynamics. We focus on the particle imbalance $I(t) = |n_\mathrm{e}(t) - n_\mathrm{o}(t)|/[n_\mathrm{e}(t)+n_\mathrm{o}(t)]$ as measured in experiments on isolated systems \cite{schreiber2015observation}. Here $n_\mathrm{e}(t)$ and $n_\mathrm{o}(t)$ are the occupations of the even and odd lattice sites, respectively. In the case of an initial density wave $\ket{\Psi(0)}=\ket{0101\ldots}$ the imbalance can be rewritten as \begin{equation}
	\label{eq:imbalance}
	I(t) = \frac{1}{N}\left| \sum_{i=1}^{L} (-1)^i \braket{\Psi(t)|\hat{n}_i|\Psi(t)} \right|,
\end{equation} where $N=L/2$ is the total number of particles. The state of the system evolves according to \begin{equation}
    \label{eq:postselection-time-evo}
	\ket{\Psi(t)}=\frac{\exp(-\mathrm{i} \hat{H} t) \ket{\Psi(0)}}{||\exp(-\mathrm{i} \hat{H} t) \ket{\Psi(0)}||}
\end{equation} where the normalization is explicitly enforced; the operator $\exp(-\mathrm{i}\hat{H}t)$ does not preserve the norm of $\ket{\Psi(0)}$ when $\hat{H}$ is non-Hermitian. In the context of Lindbladian dynamics this corresponds to trajectories which are post-selected on the absence of quantum jumps \cite{daley2014quantum,ashida2020non}.

\begin{figure}[t!]
	\centering
	\includegraphics[width=3.2in]{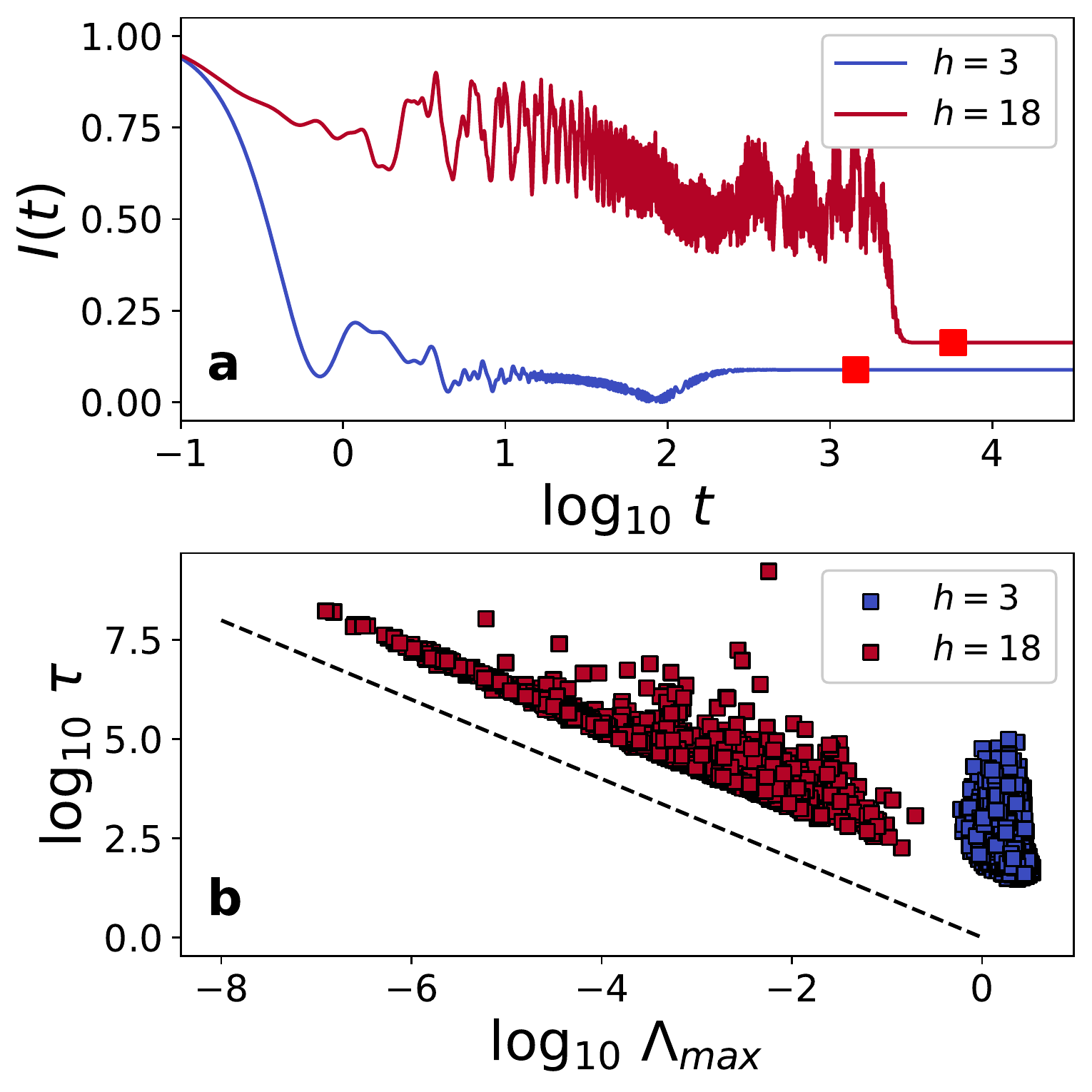}
	\caption{
		\textbf{The clustering of the relaxation timescale $\tau$ of the particle imbalance $I(t)$ is controlled by the largest imaginary eigenvalue $\Lambda_{max}$ at strong disorder.} \textbf{a} Time-evolution of $I(t)$ for a fixed realization of the disorder with $h=3$ (blue) and $h=18$ (burgundy) selected to have at least one complex eigenvalue pair in the spectrum. We set the non-Hermiticity parameter to $g=0.5$ and the system size is $L=14$. $I(t)$ is well approximated by a steady state after a time $\tau$ as indicated by the red squares. We define $\tau$ as the time where $|I(t)-I(\infty)|/|I(\infty)|<\epsilon$ for all $t>\tau$, where $\epsilon=10^{-8}$ and $I(\infty) = I(t\to\infty)$. The latter is inferred from exact diagonalization by setting $\ket{\Psi(t)}$ in Eq.~(\ref{eq:imbalance}) as the right eigenstate with the largest imaginary eigenvalue. \textbf{b} Distribution of $\tau$ for different realizations of the disorder with $h=3$ (blue) and $h=18$ (burgundy). In the localized phase the distribution of timescales clusters around a peak maximum at $\tau = \alpha \Lambda_\mathrm{max}^{-1}$, where $\log_{10}\alpha \approx 1.3$ depends on the definition of $\tau$.
	}
	\label{fig:tss_scatter}
\end{figure}

To explore the impact of complex eigenvalues on the relaxation dynamics, we consider the formation of a steady state in $I(t)$. In Fig.~\ref{fig:tss_scatter}(a) we plot the time-evolution of $I(t)$ in both the delocalized and localized regimes corresponding to $h=3$ and $h=18$, respectively. In each case we take a single realization of the disorder with at least one complex conjugate eigenvalue pair in the spectrum. It is readily seen that $I(t)$ approaches a steady state after a time $\tau$, as indicated by the red square markers. Heuristically, one expects that $\tau$ is governed by the largest imaginary eigenvalue $\Lambda_{max}$, corresponding to the largest rate of amplification; the positive and negative imaginary parts of the complex eigenvalue pairs correspond to amplification and decay, respectively. This is borne out in Fig.~\ref{fig:tss_scatter}(b) which shows the distribution of $\tau$ for different disorder realizations and two values of the disorder strength $h$; only realizations with complex eigenvalues in their spectra are considered for this analysis. The dashed line is a guide to the eye showing $\tau = \Lambda_\mathrm{max}^{-1}$. A notable feature of Fig.~\ref{fig:tss_scatter}(b) is that the distribution of timescales changes markedly in going from the weak to the strong disorder regime. In the localized phase the distribution of the scaled quantity $\tau\Lambda_\mathrm{max}$ shows a concentration of timescales in the vicinity of $\tau = \alpha \Lambda_\mathrm{max}^{-1}$, where $\log_{10}\alpha \approx 1.3$. However, in the weak disorder regime we see a broader distribution of timescales which are only bounded from below by $\Lambda_\mathrm{max}^{-1}$. Moreover, the distribution of timescales becomes elongated along the axis $\tau = \alpha \Lambda_\mathrm{max}^{-1}$ in the vicinity of the transition region (region II), as depicted in Figs~\ref{fig:g-vs-h} and \ref{fig:u-vs-h}.

\begin{figure}[t!]
	\centering
	\includegraphics[width=3.2in]{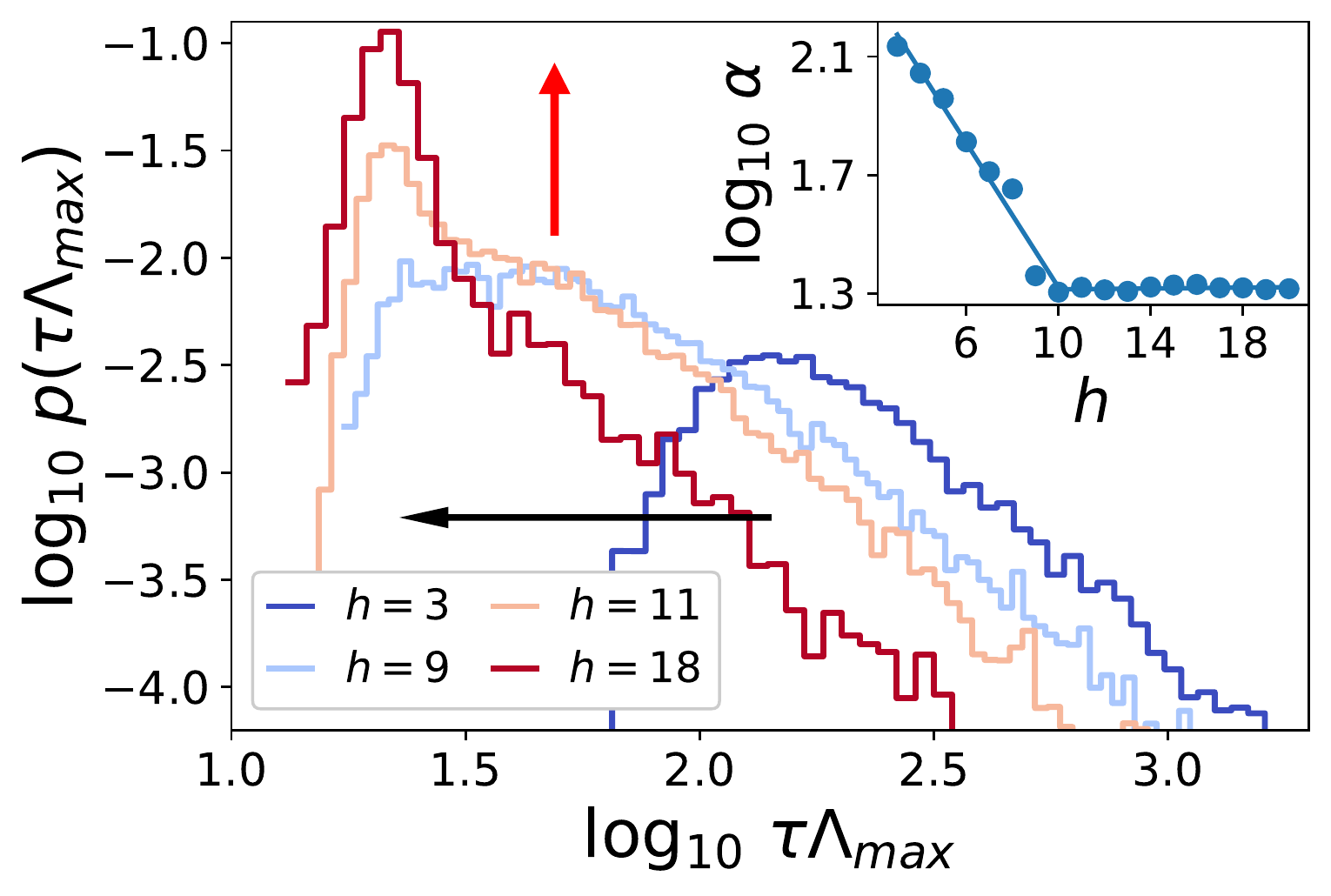}
	\caption{
		\textbf{The distribution $\tau \Lambda_\mathrm{max}$ develops a sharp peak near region II of the phase diagram.} Distribution of $\tau \Lambda_\mathrm{max}$ as a function of the disorder strength $h$. The non-Hermiticity parameter is set to $g=0.5$ and the system size is $L=14$, as used in Fig.~\ref{fig:tss_scatter}. The distribution exhibits a broad maximum which shifts towards lower values with increasing disorder strength $h$, as indicated by the black arrow. In the vicinity of the transition region (region II) shown in Figs~\ref{fig:g-vs-h} and \ref{fig:u-vs-h} the distribution develops a sharp peak, as indicated by the red arrow. The location of this peak occurs at $\tau=\alpha\Lambda_\mathrm{max}^{-1}$, with $\log_{10}\alpha \approx 1.3$ in agreement with Fig.~\ref{fig:tss_scatter}. Inset: The location of the overall broad peak for $L=14$ moves to lower values of $\alpha = \tau \Lambda_\mathrm{max}$ with increasing disorder strength. The value of $\alpha$ becomes independent of $h$ in the localized phase with $\log_{10}\alpha \approx 1.3$.
	}
	\label{fig:tss_distr}
\end{figure}

To explore this further we plot the distribution of $\tau \Lambda_\mathrm{max}$ for different values of the disorder strength. The distribution develops a sharp peak in the vicinity of the eigenstate transition at $h \approx 10$, for the chosen parameters. The peak location at $\alpha=\tau\Lambda_\mathrm{max}$ with $\log_{10} \alpha \approx 1.3$ coincides with the relationship observed in Fig.~\ref{fig:tss_scatter}, and does not change with increasing disorder strength. To show this more clearly, in the inset of Fig.~\ref{fig:tss_distr} we plot the evolution of $\tau\Lambda_\mathrm{max}$ at the overall peak of the distribution shown in the main panel. The peak location shifts to lower values with increasing $h$, becoming independent of $h$ in the localized phase. This is consistent with a direct relationship between $\tau$ and $\Lambda_\mathrm{max}^{-1}$ in the localized phase. This is in marked contrast with the thermal phase which shows a broader distribution without a sharp peak.

\section*{Discussion}\label{sec12}

The preceding analysis shows that the phase diagram in Figs~\ref{fig:g-vs-h} and \ref{fig:u-vs-h} can be interpreted in terms of the dynamical behavior of physical observables. In particular, the presence of complex eigenvalues directly impacts on the memory lifetime of the initial state. For weak disorder (region I), this memory is lost due to the presence of complex eigenvalues, reflecting delocalized eigenstates and non-Hermiticity. For intermediate disorder (region II), the memory of generic initial states is still eroded due to complex eigenvalues, despite the onset of localization. Nonetheless, the presence of localized real eigenstates can lead to long-lived memory of the initial conditions. Only at strong disorder (region III), where the vast majority of the eigenvalues are localized, and the eigenvalues are real, do generic initial states have long lifetimes. However, even here, this lifetime is not infinite - there is always some residual decay due to the presence of complex eigenvalues.

Although we cannot exclude the possibility that the eigenstate and spectral transitions merge in the thermodynamic limit, the crossing points are well separated for a broad range of parameters and accessible system sizes. This suggests the possibility of an intermediate region for at least part of the parameter space. In contrast, in the single-particle limit of the model \eqref{eq:mbl-nh-model}, delocalization is always accompanied by complex eigenvalue formation and \textit{vice versa}. As such, an intermediate region is absent in the single-particle case.

\section*{Conclusion}

In this work we have investigated the phase diagram of the interacting Hatano-Nelson model as a function of the interaction strength and the non-Hermiticity. We have mapped out two regimes for MBL \textit{via} the mid-spectrum eigenstates. In particular, we have shown that the delocalization instability of the real eigenstates occurs at a weaker disorder strength than the transition to a predominantly real spectrum. We have also explored the non-equilibrium dynamics of this model and shown the appearance of a dynamical signature in the vicinity of the eigenstate transition. It would be interesting to explore the phase diagram of this problem in the framework of Lindbladian dynamics. Finally, we note that the apparent extrapolation of region II to infinitesimal non-Hermiticity in Fig. \ref{fig:g-vs-h} is suggestive of it probing the separation between landmarks in the Hermitian MBL transition \cite{morningstar2022avalanches}. To rigorously establish or exclude such a connection would be an interesting subject of future research.

\section*{Methods}

The eigenvalues and eigenvectors of the model \eqref{eq:mbl-nh-model} are calculated \textit{via} an equivalent spin model using ED. The hard-core bosons are mapped to spins using the transformation $\hat{b}_i \to \hat{S}_i^-$, $\hat{b}_i^\dagger \to \hat{S}_i^+$, $\hat{b}_i^\dagger \hat{b}_i \to \hat{S}_i^z+1/2$, where $\hat{S}_j^\pm = \hat{S}_j^x \pm \hat{S}_j^y$ are raising and lowering operators and $\hat{S}_j^\alpha$ is the $\alpha$-projection of the spin. The spin Hamiltonian is \begin{equation}
    \begin{aligned}
        \hat{H}=\sum_{i=1}^{L}&\left[-J\left(\mathrm{e}^{-g} \hat{S}_{i}^\mathrm{-} \hat{S}_{i+1}^\mathrm{+} + \mathrm{e}^{g} \hat{S}_{i}^\mathrm{+} \hat{S}_{i+1}^\mathrm{-}\right) + U \left(\hat{S}_{i}^\mathrm{z}+1/2\right) \left(\hat{S}_{i+1}^\mathrm{z}+1/2\right) \right. \\&+ \left. h_{i} \left(\hat{S}_{i}^\mathrm{z}+1/2\right)\right].
    \end{aligned}
\end{equation} The matrix elements of $\hat{H}$ are computed in a basis of product states using QuSpin \cite{weinberg2017quspin}. Figs~\ref{fig:g-vs-h}-\ref{fig:u-vs-h} are obtained from the midspectrum eigenstates and eigenvectors using the shift-invert ED routine in SciPy \cite{virtanen2020scipy}; related algorithms have been used in the Hermitian case \cite{luitz2015many,pietracaprina2018shift}. The right eigenvectors are obtained directly, and the left eigenvectors are obtained from the right eigenvectors of the transposed matrix. 
The dynamics in Figs \ref{fig:tss_scatter} and \ref{fig:tss_distr} is obtained by iterating Eq. \eqref{eq:postselection-time-evo} using small time steps $\delta t$. This avoids the growth of the norm due to complex eigenvalues. Explicitly, we decompose $\exp(-i\hat{H}\delta t)=\sum_{k=1}^\mathcal{N} \ket{E_k}_\mathrm{R} \exp(-iE_k \delta t) \leftindex_{\mathrm{L}}{\bra{E_k}}$ using the completeness relation $\hat{I} = \sum_{k=1}^\mathcal{N} \ket{E_k}_\mathrm{R}\leftindex_{\mathrm{L}}{\bra{E_k}}$  \cite{ashida2020non}, where $\mathcal{N}$ is the dimension of the Hilbert space. The largest time step $\delta t$ is chosen so that the norm of a vector with components $\exp(-i E_k \delta t)$ is kept below $10^{-10}$. The data in Fig.~\ref{fig:tss_scatter}(a) is obtained by sampling on a linear grid until $t=10^{-1}$ and logarithmically thereafter.

\backmatter

\section*{Declarations}

\bmhead{Data availability} The simulation  data that have been generated and analyzed during this study are deposited in the King's Open Research Data System (\href{https://doi.org/10.18742/25130666}{DOI:10.18742/25130666}) \cite{reserach_data}.

\bmhead{Acknowledgments}

We acknowledge helpful discussions with Jonas Richter. J.M. acknowledges support from the EPSRC CDT in Cross-Disciplinary Approaches to Non-Equilibrium Systems (CANES) \textit{via} [grant number EP/L015854/1]. M.J.B. acknowledges support of the London Mathematical Laboratory. A.P. was funded by the European research Council (ERC) under the European Union’s Horizon 2020 research and innovation program \textit{via} Grant Agreement No. 853368. We are grateful to the UK Materials and Molecular Modelling Hub for computational resources, which is partially funded by EPSRC [EP/P020194/1 and EP/T022213/1].

\bmhead{Author contributions} A.P. and M.J.B. conceived and supervised the project. J.M. performed all of the numerical simulations. All authors analyzed the results and contributed to the writing of the manuscript.

\bmhead{Competing interests} The authors declare no competing interests.


\clearpage

\end{document}


\title[Supplementary Information for Statics and Dynamics of non-Hermitian Many-Body Localization]{Supplementary Information for Statics and Dynamics of non-Hermitian Many-Body Localization}


\author*[1]{\fnm{J\'ozsef} \sur{M\'ak}}\email{jozsef.mak@kcl.ac.uk}

\author[1]{\fnm{M. J.} \sur{Bhaseen}}%

\author[2]{\fnm{Arijeet} \sur{Pal}}%

\affil[1]{\orgdiv{Department of Physics}, \orgname{King's College London}, \orgaddress{\street{Strand}, \city{London}, \postcode{ } \country{WC2R 2LS, United Kingdom}}}

\affil[2]{\orgdiv{Department of Physics and Astronomy}, \orgname{University College London}, \orgaddress{\street{Gower Street}, \city{London}, \postcode{WC1E 6BT}, \country{United Kingdom}}}

\maketitle

\section{Supplementary Note 1: The Hatano-Nelson Model in the Context of the Lindblad Equation}

To derive the model (1) in the main text in the context of open quantum systems, we first briefly revisit two equivalent descriptions of quantum systems coupled to a large, Markovian environment. The time evolution of such open systems is governed by the Lindblad equation \cite{breuer2002theory} \begin{equation}
    \label{eq:Lindblad}
    \frac{\mathrm{d}}{\mathrm{d}t}\hat{\rho}(t) = -i \left[\hat{H},\hat{\rho}(t) \right] + \sum_{\mu}\hat{L}_\mu\hat{\rho}(t)\hat{L}_\mu^\dagger - \frac{1}{2} \hat{L}_\mu^\dagger \hat{L}_\mu \hat{\rho}(t) - \frac{1}{2} \hat{\rho}(t) \hat{L}_\mu^\dagger \hat{L}_\mu.
\end{equation} The density matrix of the system is $\hat{\rho}(t)$, the closed system Hamiltonian is $\hat{H}$ and the Lindblad jump operators $\hat{L}_\mu$ describe the effect of the environment. Equivalent to the Lindblad formalism, the time evolution of open quantum systems can be formulated in terms of quantum trajectories \cite{breuer2002theory,daley2014quantum}: along a single trajectory the system is in a pure state $\ket{\Psi(t)}$ which is subject to non-Hermitian evolution between quantum jump events occurring at random times. Quantum jumps correspond to the exchange of excitations with the environment, and if no such events happen in a time interval $[t_0,t_0+t]$, the state of the system $\ket{\Psi(t)}$ evolves according to \begin{equation}
    \label{eq:postselection-time-evo}
    \ket{\Psi(t_0+t)} = \frac{e^{-\mathrm{i}\hat{H}_\mathrm{eff}t}\ket{\Psi(t_0)}}{\left|\left|e^{-\mathrm{i}\hat{H}_\mathrm{eff}t}\ket{\Psi(t_0)}\right|\right|}.
\end{equation} Here \begin{equation}
    \hat{H}_\mathrm{eff} = \hat{H} - \frac{\mathrm{i}}{2} \sum_\mu \hat{L}^\dagger_\mu \hat{L}_\mu
\end{equation} is an effective non-Hermitian Hamiltonian. The average over the stochastically evolving trajectories yields results consistent with the Lindblad equation, that is, $\hat{\rho}(t)=\overline{\ket{\Psi(t)}\bra{\Psi(t)}}$, where the overbar denotes the average over trajectories. The expectation values of observables $\hat{O}$ are given by $\mathrm{Tr}[\hat{\rho}(t)\hat{O}]=\overline{\braket{\Psi(t)|\hat{O}|\Psi(t)}}$. Whereas usually the average over trajectories is measured in experiment, in continuously monitored quantum systems individual trajectories become accessible \cite{murch2013observing,daley2014quantum}. Moreover, groundbreaking experiments have demonstrated both quantum jumps and the intermittent non-Hermitian evolution; see Refs \cite{bergquist1986observation,naghiloo2019quantum,minev2019catch}.

Consider now a closed system Hamiltonian \begin{equation}
    \hat{H} = \sum_{j=1}^L -J\cosh(g)\left(\hat{b}_{j+1}^\dagger\hat{b}_j + \hat{b}_j^\dagger\hat{b}_{j+1}\right) + U\hat{n}_j\hat{n}_{j+1} + h_j\hat{n_j}
\end{equation} and Lindblad operators \begin{equation}
    \label{eq:lindblad-jump-ops}
    \hat{L}_j = \sqrt{2J\sinh(g)} \left(\hat{b}_j - \mathrm{i}\hat{b}_{j+1}\right).
\end{equation} The symbols are defined in the main text. The effective non-Hermitian Hamiltonian becomes \begin{equation}
    \label{eq:h_eff}
    \hat{H}_\mathrm{eff} = \sum_{j=1}^{L}\left[-J\left(\mathrm{e}^{-g} \hat{b}_{j+1}^{\dagger} \hat{b}_{j}+\mathrm{e}^{g} \hat{b}_{j}^{\dagger} \hat{b}_{j+1}\right)+U \hat{n}_{j} \hat{n}_{j+1}+h_{j} \hat{n}_{j}\right] - 2\mathrm{i}J\sinh(g)\hat{N},
\end{equation} where $\hat{N}=\sum_{j=1}^L \hat{n}_j$ is the total particle number operator. This is the model (1) in the main text; the last term amounts to a constant shift, as $\hat{H}_\mathrm{eff}$ conserves the total particle number.

The effective non-Hermitian Hamiltonian \eqref{eq:h_eff} governs the time evolution of jump-free trajectories of an open system with Lindblad jump operators \eqref{eq:lindblad-jump-ops}. The realization of these jump operators in a cold atomic setup is discussed in Appendix F of Ref. \cite{gong2018topological}, and Refs \cite{lee2014heralded,mu2020emergent} discuss the implementation of related interacting non-Hermitian models. Non-Hermitian evolution has been realized with solid state qubits \cite{naghiloo2019quantum,wu2019observation} and with ultracold atoms \cite{li2019observation}. See also Ref. \cite{liang2022dynamic}, which demonstrates non-reciprocal transport on a lattice in a cold atomic system; this is an analogous situation to the one described by the model \eqref{eq:h_eff}. Recently, a non-Hermitian many-body system has been simulated on a trapped ion quantum computer \cite{noel2022measurement}. Whereas the post-selection problem hinders the implementation of thermodynamically large interacting non-Hermitian models, these experimental developments and theoretical works point towards the potential realizability of the model \eqref{eq:h_eff}, at least for moderately sized systems.

\renewcommand\refname{Supplementary References}